\documentclass[aps,pra,10pt,twocolumn,floatfix,nofootinbib,superscriptaddress,nobibnotes,groupedaddress]{revtex4-2}

\usepackage{amsmath,amssymb,bm}
\usepackage[colorlinks=true,citecolor=cyan,linkcolor=magenta,filecolor=magenta]{hyperref}
\usepackage{graphicx}
\usepackage{microtype}
\usepackage{times}
\usepackage{comment}
\usepackage{orcidlink}
\usepackage{mathrsfs}
\usepackage[normalem]{ulem}         

\def\bra#1{\left<{#1}\right|}	
\def\ket#1{\left|{#1}\right>}	
\def\braket#1#2{\left<{#1}|{#2}\right>}	
\def\abs#1{\left|{#1}\right|}      

\definecolor{edit}{rgb}{0.0,0.0,0.8}

\definecolor{remove}{rgb}{0.8,0.0,0.0}

\begin{document}


\title{From quantum geometry to non-linear optics and gerbes:\texorpdfstring{\\}{} Recent advances in topological band theory}

\author{Tom\'a\v{s} Bzdu\v{s}ek\,\orcidlink{0000-0001-6904-5264}}\email{tomas.bzdusek@uzh.ch}
\affiliation{Department of Physics, University of Z\"urich, Winterthurerstrasse 190, 8057 Z\"urich, Switzerland}

\date{\today}

\begin{abstract}
Topological principles constitute at present an integral component of condensed matter physics, permeating the modern characterization of electronic states while also guiding materials design.
In this brief \emph{Perspective}, I highlight three research threads in single-particle topological band theory that have recently gained momentum: 
(\emph{i})~the rise of the \emph{quantum geometric tensor}, whose components can at present be directly accessed with optical probes; 
(\emph{ii})~the notions of \emph{delicate and multigap topology}, which fall outside the scope of tenfold way and symmetry-based indicators yet leave robust physical fingerprints; 
and (\emph{iii})~the consideration of \emph{bundle gerbes}, which capture formerly overlooked higher-form topological aspects of energy bands. 
These distinct directions have been elegantly woven together: 
delicate and multigap topological insulators have peculiar features in quantum geometry that can be conveniently captured by bundle gerbes. 
This viewpoint exposes the recently identified quantization of a non-linear optical response, and it invites deeper and systematic investigations into geometric and topological aspects of band structures beyond conventional Berryology.
\end{abstract}

\maketitle


\section{Quantum geometric tensor}

\begin{figure*}[t!]
\centering
\includegraphics[width=0.99\textwidth]{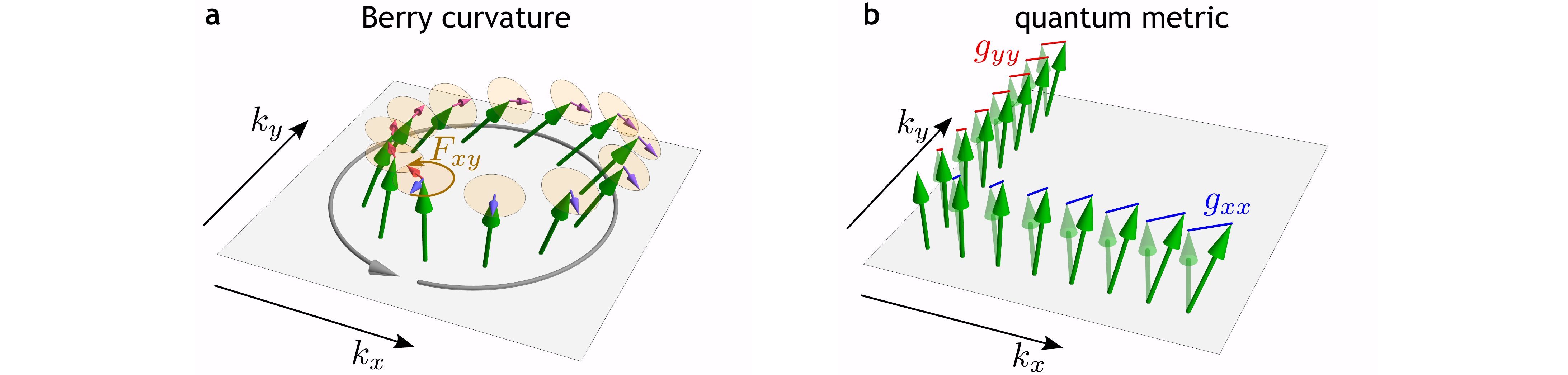}
\caption{{
\textbf{Elements of the quantum geometric tensor.} 
(\textbf{a})~Given an evolution of a Bloch state $\ket{u_n}$ (state vector, represented by green arrows) on a closed loop (circular gray arrow) inside the momentum space (plane of $k_x$ and $k_y$), Berry curvature ($F_{xy}$) quantifies the complex phase (blue-to-red arrows within the attached disks) accumulated by the state. 
(\textbf{b})~Quantum metric (here only components $g_{xx}$ and $g_{yy}$ explicitly considered) quantifies the rate of change (blue and red line segments) of the Bloch state upon variation of the momentum coordinates.
}}
\label{fig:QGT}
\end{figure*}

While \emph{Berry curvature} of Bloch states has become an established concept in condensed matter physics, entering the description of band topology~\cite{Asboth:1994} as well as semiclassical transport~\cite{Xiao:2010}, recent years have witnessed a revival of interest in a further geometric aspect of energy bands: their \emph{quantum} (or \emph{Fubini-Study}) \emph{metric}. 
Both quantities measure aspects of how the wave function changes when varying momentum.
Specifically, Berry curvature quantifies the complex phase acquired by the state when it is evolved on a small closed loop in momentum space [Fig.~\ref{fig:QGT}(\textbf{a})]; in contrast, quantum metric measures how quickly the state grows in orthogonal directions when departing away from an initial momentum [Fig.~\ref{fig:QGT}(\textbf{b})].

Both Berry curvature~\cite{Berry:1984} and quantum metric~\cite{Provost:1980} can be extracted as components of a single mathematical object: the \emph{quantum geometric tensor} (QGT).
Assuming a single non-degenerate band $\ket{u_n}$ that depends on momentum $k$, QGT is defined as the gauge-invariant Hermitian matrix~\cite{Zanardi:2007}
\begin{equation}
\label{eqn:QGT-def}
Q_{ab}^{n} = \bra{\partial_a u_n} \left(\mathbf{1} - \ket{u_n}\!\!\bra{u_n}\right)\ket{\partial_b u_n},
\end{equation}
where $\partial_a$ indicates the derivative of the state with respect to the momentum component $k_a$. 
In two and three spatial dimensions (i.e., 2D and 3D), this amounts to a $2\,{\times}\,2$ resp.~$3
\times 3$ matrix with complex entries.
The QGT naturally decomposes in the following way~\cite{Ma:2010}:
its symmetric real part is the quantum metric ($g_{ab}^n$) and its antisymmetric imaginary part is proportional to the Berry curvature ($\frac{i}{2}F_{ab}^n$) of the band $\ket{u_n}$.

Although QGT has been recognized as a property of Bloch energy bands for decades, it has recently moved into the spotlight due to the relations of its components to optical responses (discussed below), to electron-phonon coupling~\cite{Yu:2024}, and to superfluid weight in flat-band superconductors~\cite{Peotta:2015,Arbeitman:2022,Torma:2023}.
Importantly, while the QGT directly enters responses of electronic systems, the concept is equally well defined in any condensed-matter system with tunable parameters, including synthetic (and often classical) tight-binding simulators~\cite{Mao:2018,Ozawa:2019,Xue:2022,McClarty:2022,Kavokin:2022,Sahin:2025}.
For example, in the context of band theory, Ref.~\citenum{Marzari:1997} established a relation between the quantum metric and the spread functional of maximally localized Wannier~functions, which applies to both electronic and metamaterial setups.
Furthermore, quantum geometry has also been investigated in the context of Landau levels~\cite{Rhim:2020,Hwang:2021}.

Berry curvature and quantum metric are not entirely independent. 
Notably in two dimensions, where Berry curvature has only a single component $F_{12}\equiv F$, this component is bounded in absolute value by the determinant of the quantum metric; namely, 
\begin{equation}
\label{eqn:Berry-metric-inequality}
\abs{F} \leq 2 \sqrt{\det {g}}
\end{equation}
holds at every momentum $k$~\cite{Ozawa:2021}.
Correspondingly, a suitably defined (and non-quantized) integral of the quantum metric trace, called the \emph{quantum weight} ($K$)~\cite{Onishi:2024}, is bounded to be larger than the quantized integral of the Berry curvature, called the \emph{Chern number}~($C$).

The inequality~(\ref{eqn:Berry-metric-inequality}) is easily derived with a bit of algebra from the positivity of QGT. 
Specifically, let us define the rectangular matrix $A_n$ whose components $(A_n)_{\ell b} = \braket{u_\ell}{\partial_b u_n}$ correspond (up to an overal factor of $i$) to interband Berry connections. 
Since $\mathbf{1}-\ket{u_n}\!\!\bra{u_n} = P_n^\perp$ is a projector (i.e., a Hermitian matrix with eigenvalues $0$ and $1$), we can write $P_n^\perp = (P_n^\perp)^\dagger P_n^\perp$, allowing us to express the QGT as $Q^n = X_n^\dagger X_n^{\phantom{\dagger}}$ with the matrix $X_n^{\phantom{\perp}} = P_n^\perp A_n^{\phantom{\perp}}$. 
Observe that $w^\dagger Q^n w = \lVert X_n \, w\rVert^2 \geq 0$ for any choice of vector $w$, implying that eigenvalues of $Q^n$ are non-negative and, therefore, $\det Q^n \geq 0$.
On the other hand, the decomposition $Q = g + \frac{i}{2}F$ (where we dropped the band index $n$ for simplicity) in two dimensions explicitly reads
\begin{equation}
Q = \left(\begin{array}{cc}
g_{xx}      &   g_{xy} + \tfrac{i}{2}F \\
g_{xy} - \tfrac{i}{2}F     & g_{yy}
\end{array}\right),    
\end{equation}
implying that $\det Q = \det {g} - \tfrac{1}{4}\abs{F}^2$. 
From here, the inequality~(\ref{eqn:Berry-metric-inequality}) follows trivially.\footnote{One can further verify that the bound is saturated in two-band models, decomposed into Pauli matrices $\{\tau_j\}_{j\in\{x,y,z\}}$ as $H(\boldsymbol{k}) = \boldsymbol{d}(\boldsymbol{k})\cdot \boldsymbol{\tau}$.
A straightforward calculation~\cite{Piechon:2016} reveals that if we define the unit vector $\boldsymbol{n} = \boldsymbol{d}/\lVert \boldsymbol{d}\rVert$, the quantum metric becomes $g_{ab}=\tfrac{1}{4}\partial_a \boldsymbol{n}\cdot \partial_b \boldsymbol{n}$ while the Berry curvature is $F_{ab}=\tfrac{1}{2}\boldsymbol{n}\cdot\left(\partial_a \boldsymbol{n}\times \partial_b \boldsymbol{n}\right)$. 
In two dimensions this implies the equality $\abs{F}=2\sqrt{\det g}$.}

Further and less obvious constraints follow from considering optical responses, suggesting a deeper link between topology, quantum metric, and response theory.
To illustrate a specific example, recall that the linear optical conductivity [$\sigma_{ab}(\omega)$, with $a,b\in\{x,y,z\}$] can be related using the Kubo-Greenwood formula to an integral involving the interband Berry connection $A^{nm}_a=i\left<u_n|\partial_a  u_m\right>$. 
The quantities $A_a^{nm}$ correspond (up to a constant prefactor) to matrix elements of the electric dipole operator in the Bloch basis, and they encode geometrical information about the optical transitions between the energy bands $n$ and $m$~\cite{Ahn:2022}.\footnote{More precisely, the computation of $\sigma_{ab}(\omega)$ corresponds to an integral involving $A_a^{nm}A_b^{mn}\equiv Q_{ab}^{nm}$ with $n$ the occupied and $m$ the unoccupied bands; namely,
\begin{equation}
\label{eqn:Kubo-conductivity}
\sigma_{ab}(\omega) \!=\! \frac{ie^2}{\hbar} \! \! \sum_{n\neq m}
\! \oint \frac{d^3\boldsymbol{k}}{(2\pi)^3} \frac{(f_n-f_m)(E_m - E_n)}{\hbar\omega - (E_m - E_n) + i0^+}Q^{nm}_{ab},
\end{equation}
where $f_m$ is the Fermi-Dirac function at energy $E_m$, and I dropped the explicit dependence of $Q^{nm}_{ab}$ and of the band energies $E_{n}$ and $E_{m}$ on momentum for simplicity.
The quantity $Q^{nm}_{ab}$ relates back to the quantum geometric tensor, which is easily revealed with the help of the relations $\mathbf{1}-\ket{u_n}\!\!\bra{u_n} = \sum_{m\neq n}\ket{u_m}\!\!\bra{u_m}$ and $0 = \partial_a \braket{u_n}{u_m} = \braket{\partial_a u_n}{u_m} + \braket{u_n}{\partial_a u_m}$ as follows:
\begin{eqnarray}
\sum_{m\neq n} Q^{nm}_{ab} &=& -\sum_{m\neq n}\braket{u_n}{\partial_a u_m} \braket{u_m}{\partial_b u_n}\nonumber \\
&=& \bra{\partial_a u_n} \Bigg[ \sum_{m\neq n} \ket{u_m}\bra{u_m} \Bigg] \ket{\partial_b u_n} \stackrel{\textrm{Eq.~(\ref{eqn:QGT-def})}}{=} Q_{ab}^n.
\end{eqnarray}
Note also that while the dipole elements $A^{nm}_a$ and the diagonal ($n=m$) components of $Q^{nm}_{ab}$ are gauge dependent, the off-diagonal terms $Q^{nm}_{ab}$ that appear in Eq.~(\ref{eqn:Kubo-conductivity}) are gauge invariant.}
A careful analysis of the introduced concepts has been employed to reveal a non-trivial bound on the possible size of a topological energy gap. 
Given a 2D insulating system, the resulting constraint is specified as follows: define (\emph{i})~the \emph{optical weight} $W^0$ as the integral over all frequencies of the real part of the optical conductivity $\textrm{Re}[\sigma_{xx}(\omega)]$, and (\emph{ii})~let $C$ be its Chern number. 
Then, the topological energy gap obeys~\cite{Onishi:2024}
\begin{equation}
\label{eqn:bound-on-topo-gap}
\Delta \leq \frac{4 \hbar^2 W^0}{e^2\abs{C}}.
\end{equation}
Equation~(\ref{eqn:bound-on-topo-gap}) provides a clean and model-independent bound that relates experimentally accessible properties to a topological invariant.
Following this line of reasoning~\cite{Onishi:2024,Ghosh:2024}, it is further shown that integrals of quantum metric and of Berry curvature specify negative moments of the optical conductivity: up to an overall constant of $e^2/4\hbar$, the integral of $\textrm{Re}[\sigma_{xx}(\omega)+\sigma_{yy}(\omega)]/2\omega$ is equal to the quantum weight $K$, while the integral of $\textrm{Im}[\sigma_{xy}(\omega)]/\omega$ amounts to the Chern number $C$.

\begin{figure*}[t!]
\centering
\includegraphics[width=0.995\textwidth]{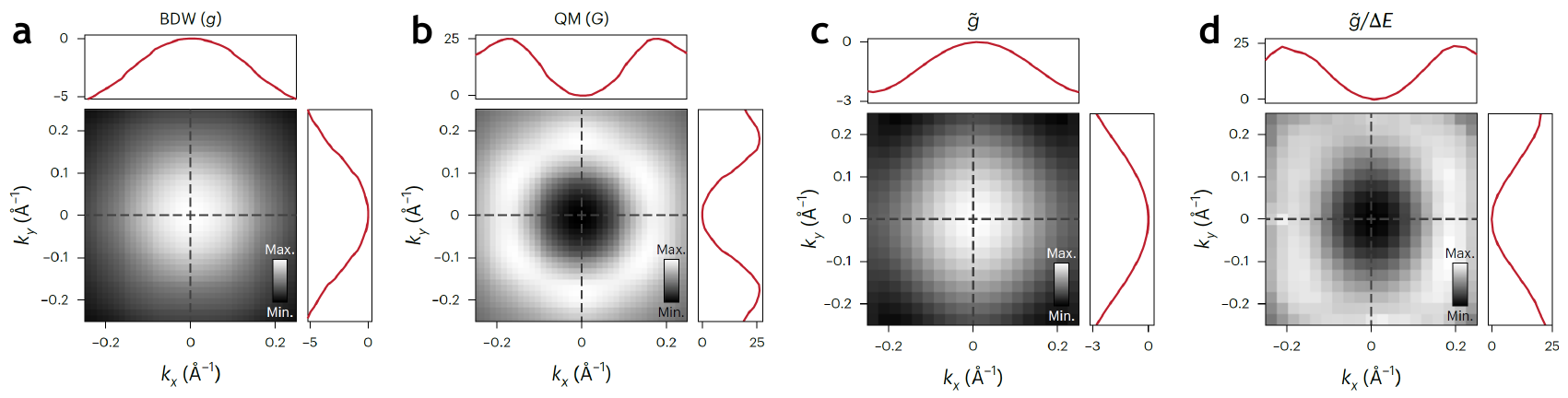}
\caption{{
\textbf{Accessing the quantum geometric tensor.} 
(\textbf{a,b}) Theoretical value of the band Drude weight and quantum metric in kagome material $\mathrm{CoSn}$, and
(\textbf{c,d}) their experimental estimates obtained through optical probes. 
(Image source:~Ref.~\citenum{Kang:2025}).
}}
\label{fig:measured}
\end{figure*}

Very recently, the relations between the components of QGT and optical responses have been turned around: the work of Ref.~\citenum{Kang:2025} proposed a protocol for mapping QGT in the momentum space of 2D materials using momentum-resolved optical probes.
Their starting point is that QGT of band $\ket{u_n}$ with energy $E_n$ can be expanded using the full eigenspectrum  of the Bloch Hamiltonian $H$ as\footnote{This result follows from Eq.~(\ref{eqn:QGT-def}) after first using $\mathbf{1} - \ket{u_n}\bra{u_n} = \sum_{m \neq n} \ket{u_m}\bra{u_m}$, leading to $Q_{ab}^n = \sum_{m\neq n}\braket{\partial_a u_n}{u_m}\braket{u_m}{\partial_b u_n}$ 
Next, by taking the derivative of $H\ket{u_n} = E_n\ket{u_n}$ we find
\begin{equation}
(\partial_a H) \ket{u_n} + H \ket{\partial_a u_n} = (\partial_a E_n) \ket{u_n} + E_n \ket{\partial_a u_n},
\end{equation}
which after projection onto $\bra{u_m}$ with $m\neq n$ results in
\begin{equation}
\bra{u_m} (\partial_a H) \ket{u_n} + E_m \braket{u_m}{\partial_a u_n} = E_n \braket{u_m}{\partial_a u_n}.   
\end{equation}
Equation~(\ref{eqn:QGT-derived}) follows from applying the derived identity $\braket{u_m}{\partial_a u_n} = \bra{u_m} (\partial_a H) \ket{u_n}/(E_n-E_m)$ twice.}
\begin{equation}
\label{eqn:QGT-derived}
Q^{n}_{ab} = \sum_{m\neq n} \frac{\bra{u_n}\partial_a H\ket{u_m}\bra{u_m}\partial_b H\ket{u_n}}{(E_m-E_n)^\alpha}    
\end{equation}
with exponent $\alpha = 2$. 
The key insight here is that the same expression with the exponent set to $\alpha = 1$, dubbed \emph{quasi-QGT} (labeled $q_{ab}^{n}$), is conveniently accessed in experiments: its real part (the \emph{band Drude weight}) can be estimated from band dispersion, accessible in conventional angle-resolved photoemission spectroscopy (ARPES), while its imaginary part (the \emph{orbital angular momentum}) is extracted with circular-dichroism ARPES, in which one measures the difference in photoemitted electron intensity under using right-handed and left-handed circularly polarized light.
Assuming an effective two-band approximation for the bands nearest to the Fermi level, the two tensors are related by simple rescaling $Q_{ab}^n=q_{ab}^n/\Delta E$, with the $k$-dependent energy splitting $\Delta E$ readable from ARPES.
The authors of Ref.~\citenum{Kang:2025} successfully illustrated their protocol with an application to the quasi-two-dimensional kagome material~CoSn, as illustrated with their data reproduced in Fig.~\ref{fig:measured}(\textbf{a}-\textbf{d}).

The outlined experimental protocol was further advanced with the subsequent work of Ref.~\citenum{Kim:2025}, which extracted the quantum metric tensor of the valence band of black phosphorus without resorting to the auxiliary quasi-QGT.
Such a simplification was achieved by reconstructing the pseudospin texture by means of polarization-resolved ARPES measurements.
Note that, beyond electronic systems, QGT is accessible by virtue of the defining relation~(\ref{eqn:QGT-def}) in synthetic setups that allow for direct measurement of the Bloch wave functions, as has been demonstrated with ultracold atoms~\cite{Yi:2023}, and when one has experimental access to the full Bloch Hamiltonian, as has been shown with photonic lattices~\cite{Guillot:2025}.

The relevance of QGT to optics does not terminate with linear (first-order) responses; rather, it extends to non-linear (higher-order) responses as well.
In such cases, the conductivity becomes a higher-rank tensor, which encodes the electric current response to the applied electric field through\footnote{In this expression, various choices of the frequency arguments and spatial indices for the current and electric fields are possible, depending on the physical mechanism and type of response (e.g., shift current, injection current, second-harmonic generation, jerk response) that one chooses to investigate.} $j_a \propto \sigma_{abc\ldots} E^b E^c \cdots$, where the coefficients $\sigma_{abc\ldots}$ are expressed\footnote{\label{foot:C-connection}For example, the shift-current conductivity $\sigma^{abc}_\textrm{shift}(\omega)$ (discussed further in Sec.~\ref{sec:deli+MG}) can be expressed as a momentum-space integral involving $C^{mn}_{cab} - (C^{mn}_{bac})^*$, where the \emph{Hermitian connection} is related to the interband Berry connection through $C^{mn}_{abc} = A_a^{nm}(\partial_b A_c^{mn})$~\cite{Ahn:2022}. In spinless-$PT$-symmetric systems, the Hermitian connection becomes real, and the linear combination $C^{mn}_{abc}-C^{mn}_{acb} = T^{mn}_{abc}$ is called interband torsion.} as integrals of combinations of dipole operators (i.e., interband Berry connections) $A_a^{nm}$ and of their derivatives~\cite{Ahn:2022,Gassner:2023,Avdoshkin:2025,Mitscherling:2025}.\footnote{See also Refs.~\citenum{Hetenyi:2023} and \citenum{Hetenyi:2022} for a complementary cumulant-based formulation of higher-order quantum geometric structures.}
Notably, recent reformulation of certain second-order responses has revealed previously unrecognized \emph{topological} aspects of band theory in three dimensions.
To illuminate this hidden topology, it is valuable to briefly turn attention to the study of few-band toy models with certain unstable topological invariants.

\section{Delicate and multigap topology}\label{sec:deli+MG}

The `stability' of a topological invariant means that it (\emph{i})~can be defined in models with an arbitrarily large number of energy bands and (\emph{ii})~cannot be trivialized through the addition of a trivial band (specifically: of an atomic limit) to either the occupied or the unoccupied sector~\cite{Kitaev:2009,Ryu:2010}.
Stable topological invariants have been extensively classified across a broad range of symmetry settings (including Altland-Zirnbauer symmetry classes~\cite{Altland:1997}, various spatial dimensions, and space groups), with the language of symmetry-based indicators~\cite{Po:2017,Kruthoff:2017} and elementary band representations~\cite{Bradlyn:2017} providing particularly valuable insights for first-principles calculations and materials design.
(So-called \emph{fragile topology}~\cite{Po:2018} corresponds to an intermediate scenario where the topological obstruction can be removed by including a trivial band to the occupied---but not by adding it to the unoccupied---part of the spectrum.)

\begin{figure*}[t!]
\centering
\includegraphics[width=0.99\textwidth]{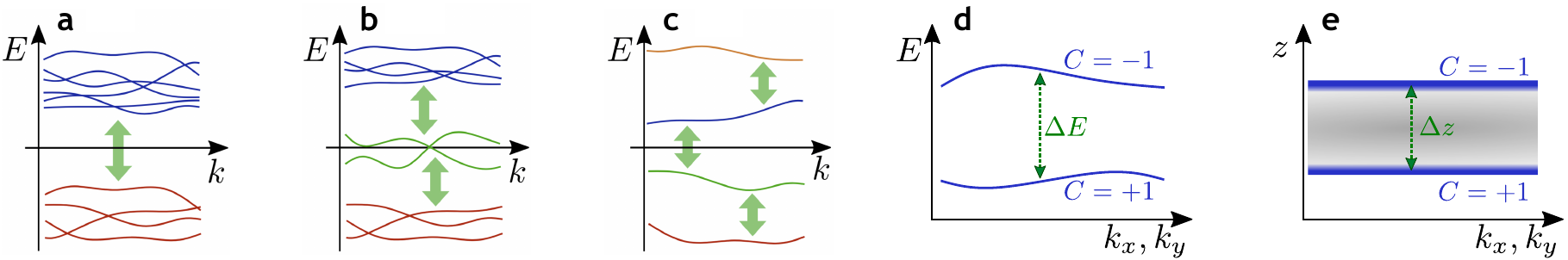}
\caption{{
\textbf{Multigap and delicate topological insulators.} 
(\textbf{a})~Conventionally, topological invariants are based on partitioning of the energy bands with a single energy gap determined by the Fermi level. 
(\textbf{b,c})~Recently, partitioning of energy bands by multiple gaps led to valuable insights into topological invariants and non-linear optical responses. 
Alternatively, if a topological invariant persists only for a small total number of energy bands, the topology is described as delicate.
(\textbf{d}) Chern insulator in two dimensions exhibits energy bands characterized by opposite Chern numbers $\pm C$ which reside in the same physical space but are separated by an energy gap ($\Delta E$).
(\textbf{e}) In constrast, Hopf insulator in a three-dimensional slab exhibits surface features with opposite Chern numbers $\pm C$ that are physically separated by the system size ($\Delta z$).
}}
\label{fig:multigap}
\end{figure*}

Nevertheless, there are instances of refined topological invariants that apply only to models with a small total number of energy bands (\emph{delicate} topology) or to models where the bands are partitioned by more than one energy gap (\emph{multigap} topology).
Numerous instances of delicate and multigap band topology have been reported; however, in contrast to stable topology, their systematic classification is still missing.

A paradigm example of delicate topology is the Hopf insulator~\cite{Moore:2008}: a two-band model of a three-dimensional magnetic topological insulator that long resisted integration into a broader classification principle.
Its simplest instance, proposed by Ref.~\citenum{Moore:2008}, corresponds to the Hamiltonian $H(\boldsymbol{k}) = \boldsymbol{d}(\boldsymbol{k})\cdot \boldsymbol{\tau}$ with components $ d_j = \zeta^\dagger \cdot \tau_j \cdot \zeta$ specified through the spinor
\begin{equation}
\label{eqn:MRW-Hopf}
\zeta(\boldsymbol{k}) = \left(\begin{array}{c}
\sin k_x + i \sin k_y \\
\sin k_z + i\left(\tfrac{3}{2}-\sum_{j=1}^3\cos k_j\right)
\end{array}\right).    
\end{equation}
The topological invariant of such two-band models is the \emph{Hopf invariant}, $h \in \mathbb{Z}$, and it is expressed elegantly as the integral of $\boldsymbol{A}\cdot \boldsymbol{F}$, i.e., of the dot product\footnote{More precisely, by interpreting the Berry connection $A$ and the Berry curvature $F$ as differential forms, the Hopf invariant is defined as $h = \frac{1}{4\pi^2}\oint d^3\boldsymbol{k} \, A \wedge F$, with the connection computed from the negative-energy Bloch state $\ket{u_-(\boldsymbol{k})}$ expressed in a \emph{global} smooth gauge.} of the Berry connection and Berry curvature of the occupied band. 
The quantization of this integral is ensured only in strictly two-band models, indicating the loss of the Hopf topology in models involving three or more energy bands.

Multigap topology [Fig.~\ref{fig:multigap}(\textbf{a}-\textbf{c})] got onto the scene with the discovery of non-Abelian braiding of band degeneracies in the momentum space of spinless-$PT$-symmetric materials~\cite{Wu:2019}, i.e., those that exhibit combined space-time-inversion symmetry and negligible spin-orbit coupling.
Specifically, while Dirac points in 2D band structures (such as that of graphene) carry a non-trivial value of a $\mathbb{Z}_2$ invariant (namely, a quantized Berry phase), whether a pair of Dirac points actually annihilate upon collision or not is dictated by their braiding around Dirac points in adjacent energy gaps.
This \emph{reciprocal braiding} of Dirac points in momentum space~\cite{Bouhon:2020}, experimentally demonstrated with acoustic meta-materials~\cite{Jiang:2021}, is captured by assigning each Dirac point a non-Abelian invariant with values in the algebra of Pauli operators $\{\pm \mathbf{1},\pm i \tau_x,\pm i\tau_y,\pm i \tau_y\}$: Dirac points in one energy gap carry `charge' $\pm i\tau_x$ while those in an adjacent energy gap carry `charge' $\pm i \tau_y$. 
The non-trivial braiding follows from the non-commutativity $\tau_x \tau_y = - \tau_y \tau_x$. 
By interpreting the time-direction as a third momentum component, the reciprocal braiding has revealed previously unknown constraints on allowed arrangements of nodal lines~\cite{Tiwari:2020} and triple nodal points~\cite{Lenggenhager:2021} in spinless-$PT$-symmetric systems.

An example of multigap band topology that is particularly easy to comprehend, although one relying on rather uncommon symmetry settings, corresponds to chiral-symmetric three-band models in three dimensions. 
Chiral operator $\Gamma$ (subject to $\Gamma^2 = \mathbf{1}$) constrains the Bloch Hamiltonian $H(\boldsymbol{k})$ through $\Gamma H(\boldsymbol{k}) \Gamma = - H(\boldsymbol{k})$, and it implies that if energy $E$ is an eigenvalue at $\boldsymbol{k}$ then so is the negative $-E$. 
In particular, a chiral-symmetric Hamiltonian with an odd number of orbitals is enforced to exhibit a flat band at $E=0$.
Choosing the chiral operator as $\Gamma = (+1,-1,-1)$, the most general Hamiltonian takes the form~\cite{Neupert:2012,Palumbo:2019}
\begin{equation}
\label{eqn:chiral-model}
H = \left(\begin{array}{ccc}
0   & d_1 + i d_2 & d_3 + i d_4 \\
d_1 - i d_2 & 0 & 0 \\
d_3 - i d_4 & 0 & 0
\end{array}\right)    
\end{equation}
and its eigenvalues are $E_0 = 0$ and $E_\pm = \pm \lVert \boldsymbol{d}\rVert$ with $\boldsymbol{d} = (d_1,d_2,d_3,d_4)$.
For the three energy bands to be separated by energy gaps, it is necessary that $\lVert \boldsymbol{d}\rVert > 0$ at all momenta, which allows us to define a unit vector $\boldsymbol{n} = \boldsymbol{d}/\lVert \boldsymbol{d} \rVert \in S^3$ that depends smoothly on momentum.
Assuming a three-dimensional model, its 3D momentum space can wind around this three-sphere.
This can be ensured through the choice $d_j = \sin k_j$ for $j={1,2,3}$ and $d_4 = M - \sum_{j=1}^3 \cos k_j$ with tunable parameter $M$.
Topological invariant of this specific example of a multigap topological insulator is the integer-valued three-dimensional winding number, expressed using the fully antisymmetric Levi-Civita symbols $\epsilon$ as
\begin{equation}
\label{eqn:3D-winding}
W_3 = \frac{1}{12\pi^2}\oint d^3 \boldsymbol{k} \epsilon^{j k \ell m}  \epsilon^{abc} n_j (\partial_a n_k) (\partial_b n_\ell) (\partial_c n_m).
\end{equation}
The specified model exhibits non-trivial values of $W_3$ for parameters $\abs{M}<3$.

\begin{figure*}[t!]
\centering
\includegraphics[width=0.68\textwidth]{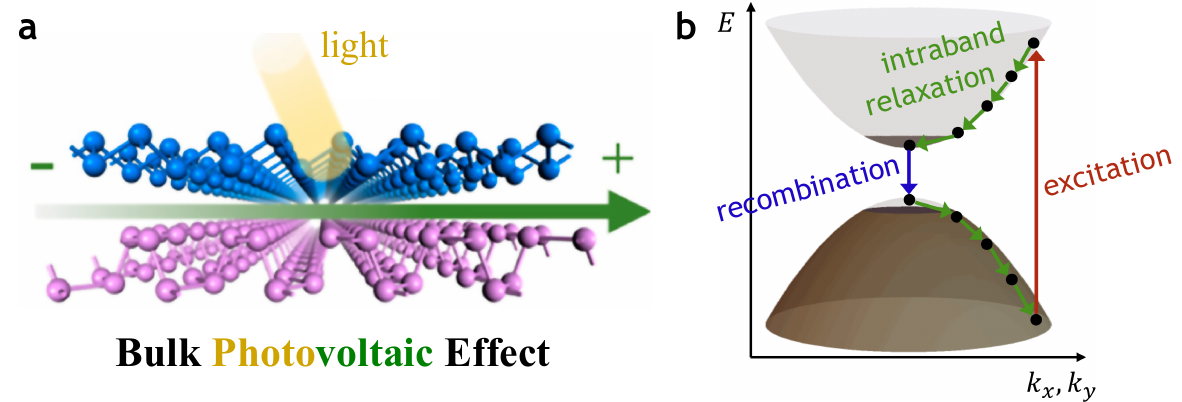}
\caption{{
\textbf{Photovoltaic shift current.} (\textbf{a}) Elementary schematic of the shift current, where a dc-current flows in a non-centrosymmetric material under illumination (without external bias).
(Image source: Ref.~\citenum{Xin:2024}.)
(\textbf{b}) Various microscopic processes contribute to the shift current. 
The discussed quantized results relate to the `excitation' component of the shift current. 
(Image adapted from Ref.~\citenum{Zhu:2024}.)
}}
\label{fig:shift}
\end{figure*}

Abundant examples of unstable invariants have been characterized over the past five years.
In few-band systems with mirror symmetry, given certain condition on mirror eigenvalues of the energy bands, \emph{returning Thouless pump} $\textrm{RTP}\in\mathbb{Z}$ invariant can be defined~\cite{Nelson:2021}, which corresponds to a quantized Chern number on half of the Brillouin zone (which is reverted on the complementary half).
Furthermore, the Hopf invariant was found to extend to models with arbitrarily many bands as long as these are all non-degenerate throughout the 3D momentum space~\cite{Lapierre:2021}. 
Crucially, spinless-$PT$-symmetric systems turned out to be a particularly fertile ground for delicate and multigap topology.
Beyond the already introduced non-Abelian invariant, any pair of energy bands in 2D can carry an \emph{Euler class} $e \in \mathbb{Z}$~\cite{Bouhon:2020}.
Non-zero values of $e$ lead to fingerprints in QGT~\cite{Jankowski:2025b} and imply that the bands must be tied together with a minimum of $2\!\abs{e}$ Dirac points.
Further topological measures (real Hopf invariants, Pontryagin invariants, and isoclinic winding numbers) were introduced for collections of bands in spinless-$PT$-symmetric models in 3D~\cite{Lim:2023,Davoyan:2024,Jankowski:2024b}.
Besides their solid theoretical understanding, multiple instances of delicate and multigap topological band structures have been realized in synthetic platforms, especially in acoustic~\cite{Jiang:2024,Cheng:2025,Mo:2025b}, optical~\cite{Guo:2021}, and electric-circuit~\cite{Wang:2023} setups.

While such unstable topological models appear to be rather remote from applications to electronic band structures in realistic crystalline solids, valuable insights have been gained by considering non-linear optical properties of such prospective electronic realizations. 
In particular, the shift current [Fig.~\ref{fig:shift}(\textbf{a})] constitutes a dc-response of non-centrosymmetric crystals to an ac-field of the form 
\begin{equation}
\label{eqn:shift-current}
j_a(0) = \sigma_{abc}^\textrm{shift}(\omega) E_b(-\omega) E_c(\omega).
\end{equation}
It was reported~\cite{Alexandradinata:2024} that models with non-trivial RTP admit a quantized contribution to the integral $\int d \omega \,  \sigma_{xyy}^\textrm{shift}(\omega)$ for a suitable orientation of the axes $x$ and $y$.
Furthermore, the integrated \emph{circular} shift photoconductivity in three dimensions is quantized if any of the broad range of unstable invariants in spinless-$PT$-symmetric setups in 3D acquires a non-trivial value~\cite{Jankowski:2024}. 
Remarkably, the mathematical descriptions developed in this context, phrased in terms of higher-form topological structures, even provide hints for transferring these quantization results towards realistic materials, as discussed further~in~Sec.~\ref{sec:outlooks}.

Three points concerning Eq.~(\ref{eqn:shift-current}) merit clarification.
First, the shift current in real materials includes multiple contributions [Fig.~\ref{fig:shift}(\textbf{b})]; namely, those induced by interband electron-hole excitations via absorption of photons ($\textrm{shift}_\textrm{exc.}$), those induced by intraband relaxation via scattering with phonons and impurities ($\textrm{shift}_\textrm{intra.}$), and those induced by recombination of electron-hole pairs ($\textrm{shift}_\textrm{rec.}$)~\cite{Zhu:2024}.
The reported quantized contribution relates solely to the `excitation' part of the shift current. 
Whether this contribution dominates the non-linear response depends on the specific material. 
Second, non-centrosymmetric crystals admit asymmetric excitation and scattering rates, enabling a further contribution to the dc-response called the ballistic current~\cite{Sturman:2020}. 
In the subsequent text, we simplify the discussion by focusing solely on the excitation part of the shift current, which has been reported to possess a topological character.
Finally, it is observed in some experiments\footnote{This is of potential concern especially when the electron correlations cannot be neglected, such as observed in the study of longitudinal circular photogalvanic current in $(\textrm{TaSe}_4)_2\textrm{I}$~\cite{Cheng:2024} or of shift current in $\textrm{ZrTe}_5$~\cite{Luo:2025}.} that an optical response can become a non-analytic function of the field fluence.
When expressing Eq.~(\ref{eqn:shift-current}), absence of such non-analyticities is implicitly assumed.

Before delving into the higher-form topological structures at play, I also mention in passing two additional idiosyncrasies of delicate and multigap topology.
First, they provide a novel paradigm for boundary anomalies. 
Recall that in conventional topological insulators one encounters `topologically interesting' energy bands (e.g., ones carrying Chern number $\pm C$) that reside in the same physical space but are separated by an energy gap into the occupied and the unoccupied sector [Fig.~\ref{fig:multigap}(\textbf{d})].
In contrast, it is found that certain unstable topological models [including the Hopf insulator in Eq.~(\ref{eqn:MRW-Hopf})] in slab geometry exhibit analogous topologically interesting bands that are instead physically separated to reside on the top vs.~the bottom surface of the slab [Fig.~\ref{fig:multigap}(\textbf{e})] (while potentially being degenerate in energy)~\cite{Alexandradinata:2021}.
The second notable aspect is related to the previous one and concerns localization on disorder. 
Strong disorder in a Hopf insulator can generate a system in which all bulk eigenstates are localized at every filling, yet the phase is topologically nontrivial and enforces delocalized boundary transport with a quantized boundary Hall conductance enabled by the surface Chern number.
Generalizing this phenomenology to higher dimensions and other Altland-Zirnbauer symmetry classes has very recently grown into the broader notion of \emph{ultra-localized topological insulators}~\cite{Lapierre:2024}.

\begin{figure*}[t!]
\centering
\includegraphics[width=0.93\textwidth]{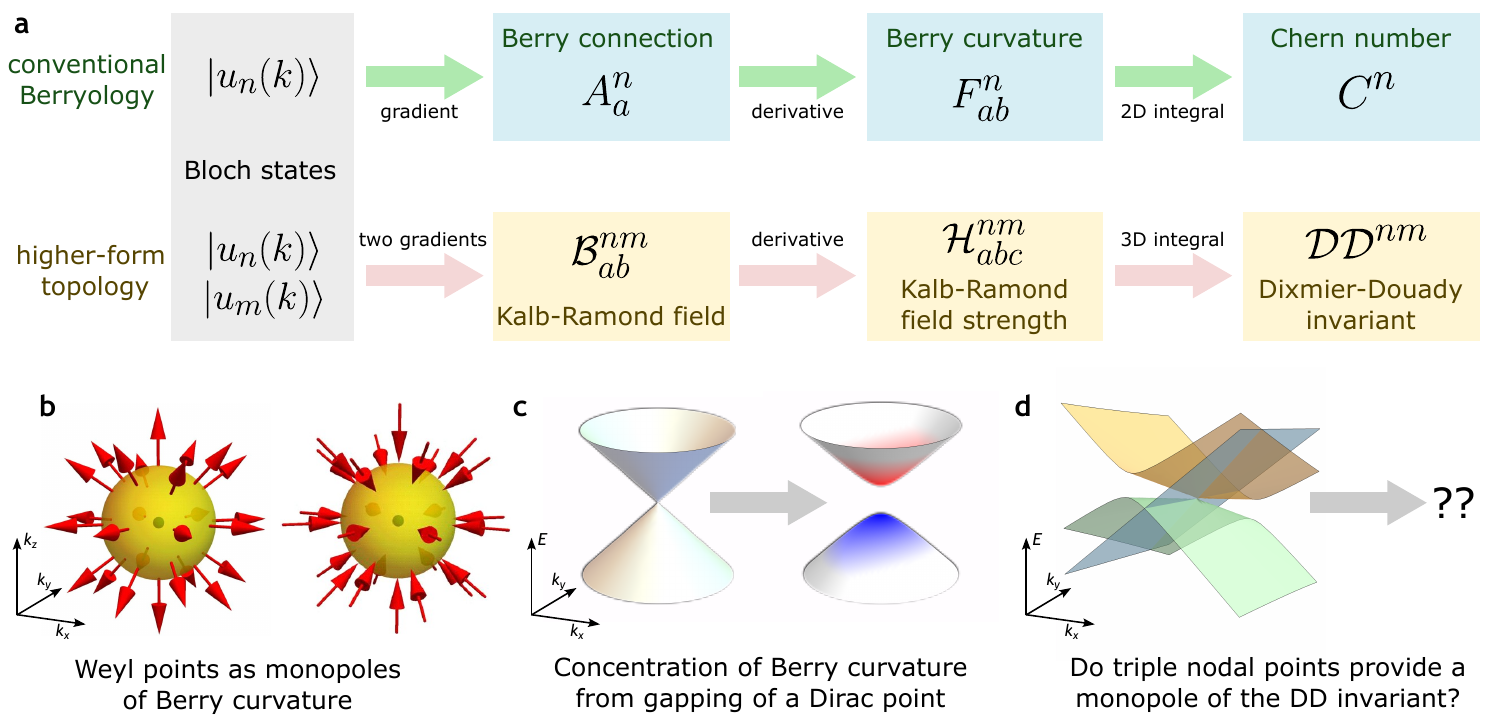}
\caption{{
\textbf{Comparing Berry quantities against higher-form topological structures.} 
(\textbf{a}) Flowchart for computing the Chern number vs.~the Dixmier-Douady ($\mathcal{DD}$) invariant.
(\textbf{b}) Weyl points (enclosed dots) in 3D crystals act as sources and drains of a quantum of Berry curvature (red arrows).
(\textbf{c}) Opening of an energy gap in Dirac points in 2D crystals generates a large concentration (specifically, a `half-qauntum') of Berry curvature.
(\textbf{d}) Shifting the focus from Berryology in 2D to higher-form topology in 3D, one is led to wonder about analogous half-quanta of the $\mathcal{DD}$ topology. 
In particular,} what are the analogous nodal features in three-band models which, after a suitable perturbation, generate quantized concentrations of the Kalb-Ramond curvature?
}
\label{fig:comparison}
\end{figure*}

\section{Higher-form topological structures}\label{eqn:higher-forms}

To introduce the higher-form topology lurking behind the quantized contributions to non-linear optical responses, it is insightful to review the key mathematical objects in the context of `Berryology' [top row of Fig.~\ref{fig:comparison}(\textbf{a})]~\cite{Vanderbilt:2018}.
The determination of Berry curvature can be understood as a three-step process: 
first, find Bloch states $\ket{u_n(k)}$ in a smooth gauge in some region of the momentum space; 
second, compute the Berry connection $A_{a}^n = i\left<u_n|\partial_a u_n\right>$ using the derivatives of the state; 
and third, compute the Berry curvature 
\begin{equation}
\label{eqn:Berry-derivative}
F_{ab}^n = \partial_a A_b^n - \partial_b A_a^n 
\end{equation}
as an antisymmetrized derivative of the connection.
If a gauge transformation $\ket{u_n(k)} \mapsto \mathrm{e}^{i \chi(k)}\ket{u_n(k)}$ is applied to the Bloch states, the Berry connection is altered, yet the Berry curvature remains unchanged.
This is mathematically equivalent to the way the scalar and the vector potential in electromagnetism are changed by gauge transformations while the electric and magnetic fields remain invariant.

\begin{figure*}[t!]
\centering
\includegraphics[width=\textwidth]{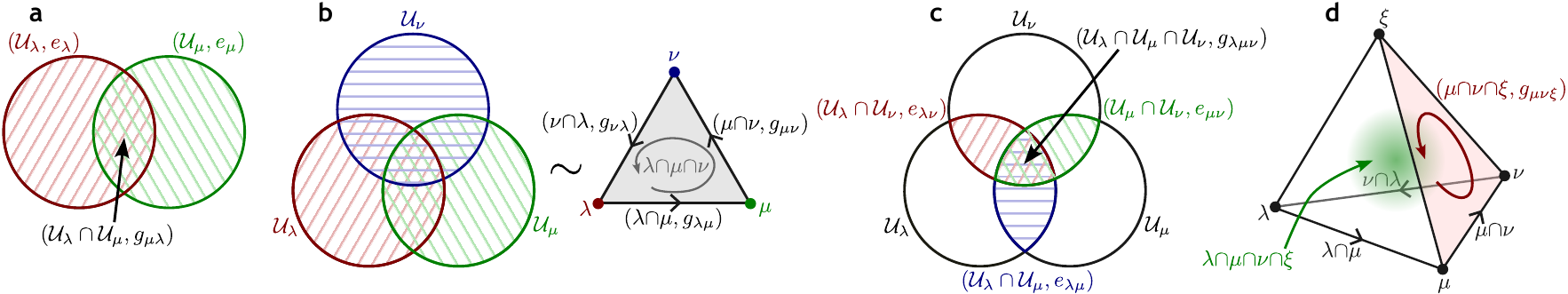}
\caption{{
\textbf{Vector bundles vs.~bundle gerbes.}
(\textbf{a})~On a vector bundle, one equips local neighborhood $\mathcal{U}_\lambda$ with basis vectors $e_\lambda$. 
On double overlaps, the two bases $e_\lambda,e_\mu$ are related by a transition function $g_{\lambda\mu}$.
(\textbf{b})~Representing a triple overlap as a triangle (whose oriented edges are double overlaps, and vertices are the individual neighborhoods), one finds that the three transition functions at the edges obey the $2$-cocycle condition.
(\textbf{c})~To obtain a bundle gerbe, we equip each double overlap $\mathcal{U}_\lambda \cap \mathcal{U}_\mu$ with a basis vector $e_{\lambda \mu}$. 
On triple overlaps, the three basis vectors are related through a generalized transition function $g_{\lambda\mu\nu}$.
(\textbf{d})~Representing a fourfold overlap as a tetrahedron, one finds that the four generalized transition functions assigned to the oriented triangular faces obey the $3$-cocycle condition.  
}}
\label{fig:transitions}
\end{figure*}

Here is how the same steps are presented in mathematical literature:
The continuous association of momenta $k$ to Bloch states $\ket{u_n(k)}$ is called a \emph{vector bundle}, since one assigns a state vector (in the Hilbert space $\mathscr{H}$) to each momentum $k$.
The notions of connection and curvature can be defined generally for any vector bundle; for example, considering vectors tangent to a two-dimensional surface results in the usual geometric (i.e., Gaussian) curvature of the surface.
The (anti-)symmetrization of the derivatives in Eq.~(\ref{eqn:Berry-derivative}) is important: it ensures that all the constructed objects are differential forms, whose integrals obey variants of Stokes' theorem~\cite{Baez:1994}.
Specifically, the Berry connection is described as a $1$-form because it involves a single derivative, while the Berry curvature constitutes a $2$-form due to the involvement of two derivatives.

The key idea behind \emph{higher-form topology} is to find analogous mathematical objects where the number of indices is increased by one [bottom row of Fig.~\ref{fig:comparison}(\textbf{a})]. 
In other words: is it possible to construct a `higher-form connection' (which involves two derivatives and that changes under gauge transformations) whose derivative gives a `higher-form curvature' (which involves three derivatives and that is gauge-invariant)?
The answer is affirmative, with the sought objects appearing in the integrated circular shift photoconductivity~\cite{Jankowski:2024b} (specifically, as alluded earlier, in the `excitation' contribution to the shift current).
The final expression (modulo unspecified coefficients and minus symmetrization over indices) takes the form~\cite{Jankowski:2025}
\begin{equation}
\label{eqn:circular-photocurrent}
\int d \omega \,\sigma_{abc}^\textrm{shift,C}(\omega) \propto \int_\textrm{BZ} d^3 k \sum_{n \in \textrm{occ.}} \sum_{m \in\textrm{unocc.}} \mathcal{H}^{nm}_{abc}.
\end{equation}
Here, the summation over all occupied and unoccupied bands on the right-hand side corresponds to integrating over all positive frequencies on the left-hand side.
While the object $\mathcal{H}^{nm}_{abc}$ on the right-hand side was originally formulated in the QGT context in terms of derivatives of the interband Berry connection~\cite{Ahn:2022}, namely as a suitably antisymmetrized interband torsion (see also footnote~\ref{foot:C-connection}), here I follow the more recent description in terms of bundle gerbes and tensor connections~\cite{Jankowski:2025}---concepts borrowed from the string theory literature~\cite{Kalb:1974}.
For simplicity, I will focus on spinless-$PT$-symmetric models. 
Because a bulk shift current generally requires broken inversion, the condition ``$PT$ present with $P$ broken'' implies $T$ is broken (for example, by antiferromagnetic order) while the combined $PT$ symmetry remains intact.

One mathematically clean way\footnote{The following two paragraphs (including Fig.~\ref{fig:transitions}) are aimed at specialists and rather technical in comparison with the rest of the text. 
They can be skipped without impacting the logic of the subsequent discussion. 
To achieve brevity, I employ the exterior derivative $d$ in the equations involved here.} of introducing bundle gerbes relates to properly generalizing the notion of gauge freedom in vector bundles~\cite{Murray:1996}.\footnote{In this exposition, I intentionally avoid the language of category theory. 
For a categorical perspective into higher gauge theory, see Refs.~\cite{Baez:2007} and~\cite{Borsten:2025}.} 
In the interesting cases where the vector bundle is topologically non-trivial (e.g., when it carries a Chern number), a \emph{global} smooth gauge for Bloch states $\ket{u_n(k)}$ over the entire Brillouin zone (BZ) does not exist.
This forces geometers to cover BZ with overlapping local neighborhoods\footnote{The specific choice of the cover is unimportant and does not affect the result as long as it is ``good''. 
This is a technical notion that means that each neighborhood, as well as every double/triple/multifold overlap, is contractible to a point.} $\mathcal{U}_\lambda$, finding a \emph{local} smooth gauge in each. 
However, when comparing the gauge on two overlapping neighborhoods, the geometers face a subtle problem: while there is a unique ray\footnote{By the `ray' of the state $\ket{\psi}$ we mean the linear span $[\psi]=\{\lambda\ket{\psi}\,|\,\lambda \in \mathbb{C}\}$, which is a linear subspace of the Hilbert space $\mathscr{H}$.} $\mathbb{C}\subset \mathscr{H}$ associated with the eigenstate $\ket{u_n(k)}$ at momentum $k\in \mathcal{U}_\lambda \cap \mathcal{U}_\mu$, this complex ray is not equipped with any canonical choice of basis vector. 
Therefore, upon comparison, the two geometers find that their basis vectors $e_{\lambda},e_{\mu}$ are generally rotated with respect to each other as $e_\lambda = g_{\lambda \mu} e_\mu$ where $g_{\lambda\mu}(k) \in U(1)$ is called a transition function [Fig.~\ref{fig:transitions}(\textbf{a})].
One can use the transition function to express the difference of the Berry connection in the two bases as $A_{(\lambda)} - A_{(\mu)} = g_{\lambda\mu}^{-1} d g_{\lambda\mu}^{\phantom{-1}}$ (as expected of a gauge transformation). 
After taking a derivative, one finds $F_{(\lambda)} = F_{(\nu)}$, i.e., the two geometers agree on the curvature.
On triple overlaps, the transition functions are shown to obey so-called $2$-cocycle condition $g_{\lambda\mu}g_{\mu\nu}g_{\nu\lambda} = 1$ [Fig.~\ref{fig:transitions}(\textbf{b})], which implies a relation to invariants in the second cohomology group $H^2(\textrm{BZ},\mathbb{Z})$.\footnote{Unsurprisingly, this invariant is precisely the Chern number. 
Upon employing Čech cohomology [replacement of $n$-fold overlaps of neighborhoods by $(n-1)$-simplices], the Chern number of the vector bundle can be extracted solely from the knowledge of the transition functions $g_{\lambda \mu}$.}

To construct a bundle gerbe, we introduce one more level of complexity: 
we attach a complex ray to all points of every \emph{double} overlap $\mathcal{U}_\lambda \cap \mathcal{U}_\mu$.
The pair of geometers that meet here agree on their shared choice of the basis vector $e_{\lambda\mu}$ for the rays on this double overlap.
Crucially, on the \emph{triple} overlap $k\in \mathcal{U}_\lambda \cap \mathcal{U}_\mu \cap \mathcal{U}_\nu$ every double overlap becomes furnished with its own basis vector, providing us with ($e_{\lambda\mu},e_{\lambda\nu},e_{\mu\nu}$).
When all three geometers meet, they compare their unit vectors using a generalized transition function $g_{\lambda\mu\nu}(k)\in U(1)$ that relates the \emph{tensor product} $e_{\lambda\mu}\otimes e_{\mu\nu} = g_{\lambda\mu\nu} e_{\lambda\nu}$ [Fig.~\ref{fig:transitions}(\textbf{c})].
One can show that the ordinary ($1$-form) connections computed from these three bases obey $A_{(\lambda\mu)} + A_{(\mu\nu)}+A_{(\nu\lambda)} = g_{\lambda\mu\nu}^{-1} d g_{\lambda\mu\nu}.$
By taking the derivative of this relation, the corresponding $2$-form curvatures obey $F_{(\lambda\mu)} + F_{(\mu\nu)} + F_{(\nu\lambda)} = 0$, which allows us to equip each neighborhood with a $2$-form potential $\mathcal{B}_{(\lambda)}$ such that $d A_{(\lambda \mu)} = F_{(\lambda\mu)}=\mathcal{B}_{(\mu)} - \mathcal{B}_{(\lambda)}$. 
Taking a derivative of the last expression and defining the 3-form $\mathcal{H}_{(\lambda)} = d \mathcal{B}_{(\lambda)}$, we find that the geometers agree on the value of $\mathcal{H}_{(\lambda)}=\mathcal{H}_{(\beta)}=\mathcal{H}_{(\gamma)}$. 
The gauge-dependent object $\mathcal{B}$ is the sought $2$-form connection (also called a \emph{tensor connection}), and the gauge-independent quantity $\mathcal{H}$ is the associated $3$-form curvature.
The presented construction also generalizes the case of vector bundles in that, on fourfold overlaps, the $3$-cocycle condition $g_{\lambda\mu\nu}^{\phantom{-1}}g_{\lambda\nu\xi}^{\phantom{-1}}g_{\mu\nu\xi}^{-1}g_{\lambda\mu\xi}^{-1}=1$ holds [Fig.~\ref{fig:transitions}(\textbf{d})], implying a relation to the third cohomology group $H^3(\textrm{BZ},\mathbb{Z})$.\footnote{
The presented construction can be extended to an arbitrary higher dimension~\cite{Palumbo:2019}. 
In crystalline solids, the dimension of the momentum space restricts the possible bundle gerbes to structures with a $2$-form connections and $3$-form curvatures; however, higher-order generalizations could, in principle, be realized in artificial setups with synthetic dimensions.}

It may appear mysterious how such a complicated mathematical object may arise in band theory; however, there is a convenient shortcut~\cite{Palumbo:2019,Guendelman:1995}. 
To proceed, one needs to identify a triplet of scalar fields $\phi_{1,2,3}$ (two complex and one ``pseudoreal'') that obey the gauge transformation 
\begin{equation}
\label{eqn:scalar-fields-GT}
\phi_1 \mapsto \phi_1 \mathrm{e}^{i \chi(k)},\; \phi_2 \mapsto \phi_2 \mathrm{e}^{-i \chi(k)},\;\textrm{and}\;\phi_3 \mapsto \phi_3 + \chi(k).
\end{equation}
With these scalar fields, one introduces the $2$-form
\begin{equation}
\label{eqn:Kalb-Ramond-B}
\mathcal{B}_{ab}^{nm} \propto \sum_{ijk} \epsilon^{ijk} \phi_i \partial_a \phi_j \partial_b \phi_k,
\end{equation}
where $\epsilon^{ijk}$ is the fully antisymmetric Levi-Civita symbol.
The field $\mathcal{B}^{nm}$, called \emph{Kalb-Ramond field}~\cite{Kalb:1974}, transforms under the change of gauge in Eq.~(\ref{eqn:scalar-fields-GT}) as
expected of a connection $2$-form.
In spinless-$PT$-symmetric models, two of these fields can be constructed from (energetically isolated) Bloch states $\ket{u_n(k)}$ and  $\ket{u_m(k)}$ upon fixing an orbital $\ket{j}$ as $\phi_1(k) = \left<j|u_n(k)\right>$ (i.e., the orbital-$j$ component of $\ket{u_n}$) and $\phi_2(k) = \left<u_m(k)|j\right>$~\cite{Jankowski:2025}. 
The remaining field $\phi_3(k)$ is defined as the line integral of the interband connection $A_{nm}$ on a suitably defined path terminating at $k$.

The three scalar fields as well as the tensor connection $\mathcal{B}$ all change under gauge transformation of the Bloch states; however, the symmetrized derivative 
\begin{equation}
\label{eqn:3-form-curvature}
\mathcal{H}_{xyz}^{nm} = \partial_x \mathcal{B}_{yz}^{nm} + \partial_y \mathcal{B}_{zx}^{nm} + \partial_z \mathcal{B}_{xy}^{nm},
\end{equation}
called the \emph{Kalb-Ramond field strength} (or sometimes \emph{higher Berry curvature}), is a gauge-invariant curvature $3$-form. 
It has been established~\cite{Jankowski:2025} that, in the presence of $PT$, the $3$-form curvature $\mathcal{H}_{xyz}^{nm}$ can be non-zero only in models with three or more energy bands.
Importantly, the integral
\begin{equation}
\mathcal{DD}^{nm} = -\frac{1}{4\pi^2} \int_\textrm{BZ} d^3k \, \mathcal{H}_{xyz}^{nm}   ,
\end{equation}
which enters the integrated circular shift photoconductivity [Eq.~(\ref{eqn:circular-photocurrent})], is an integer called the \emph{Dixmier-Douady invariant} of the bands $\ket{u_n}$ and $\ket{u_m}$, and it constitutes a higher-cohomology (and thus higher-dimensional) analog of the Chern number.
From the experimental vantage point, if the $\mathcal{DD}$ invariant is nonzero, the integrated circular shift photoconductivity changes sign under swapping the pump helicity and is stable to gentle band-structure perturbations that maintain the energy gap.
In spinless-$PT$-symmetric systems with non-degenerate bands, every pair of bands $n,m$ is characterized by a $\mathcal{DD}$ invariant (although they are not all independent of each other). 
In the presence of band degeneracies,  
quantization is only ensured if one sums contributions over all degenerate bands. 

Interestingly, it was found that the $\mathcal{DD}$ invariant is at play also for certain further unstable topological invariants beyond spinless-$PT$-symmetric systems, including the two examples explicitly discussed in Sec.~\ref{sec:deli+MG}. 
For the two-band Hopf insulator defined by Eq.~(\ref{eqn:MRW-Hopf}), the construction of the higher-form topology follows exactly the Kalb-Ramond field in Eq.~(\ref{eqn:Kalb-Ramond-B}) with the scalar fields specified therein~\cite{Jankowski:2025}.
In contrast, for the chiral-symmetric three-band model in Eq.~(\ref{eqn:chiral-model}), the same Kalb-Ramond field is utilized, but with a different choice of the scalar fields. 
Namely, expressing the negative-energy eigenstate as
\begin{equation}
\ket{u_-} = \frac{1}{\sqrt{2}} \left(-1, u_1, u_2\right)^\top 
\end{equation}
with $u_1 = (d_1-id_2)/{\lVert \boldsymbol{d} \rVert}$ and $u_2=(d_3-id_4)/{\lVert \boldsymbol{d} \rVert}$, one takes $\phi_1 = u_1$, $\phi_2 = u_1^*$, and $\phi_3 = -i \log u_2$~\cite{Palumbo:2019}. 
With this choice, one finds that the integral of the resulting $3$-form curvature reproduces the three-dimensional winding number defined in Eq.~(\ref{eqn:3D-winding}).

\section{Outlooks for future research}\label{sec:outlooks}

While topological classification of energy bands and of their bulk-boundary correspondence is at present often described as a largely solved problem~\cite{Kitaev:2009,Ryu:2010,Altland:1997,Po:2017,Kruthoff:2017,Bradlyn:2017}, recent considerations of quantum geometry and of its application to optical responses have revealed a fresh perspective where geometric and topological ideas continue to shape research into novel aspects of single-particle band theory.
These approaches have been further revamped by the recent characterization of certain delicate and multigap topological insulators, a subject of continued investigations~\cite{Moore:2008,Wu:2019,Bouhon:2020,Jiang:2021,Tiwari:2020,Lenggenhager:2021,Palumbo:2019,Nelson:2021,Lapierre:2021,Jankowski:2025b,Lim:2023,Davoyan:2024,Jankowski:2024b,Jiang:2024,Cheng:2025,Mo:2025b,Guo:2021,Wang:2023,Zhu:2024,Alexandradinata:2024,Jankowski:2024}, by means of higher-form topological structures.
The higher-form description not only presents new insights into the mathematical meaning of unstable topological invariants, but---and importantly---it provides hints of formerly overlooked geometric aspects of single-particle energy bands in general. 
Recent findings have revealed that these geometric features may, in turn, lead to observable and even quantized fingerprints in measurable response functions of crystalline compounds, thus also warranting a substantial experimental effort into electronic band structures beyond the elementary angle-resolved photoemission experiments and topological bulk-boundary correspondence.

The introduction of higher-form topological structures in the description of optical responses and in the characterization of unstable band topology invites manifold questions.
In particular, while quantum metric and tensor Berry connections have been related to each other~\cite{Palumbo:2018}, a deeper analysis still remains to be conducted.
Notably, recall that Weyl points in 3D act as monopole sources of Berry curvature [Fig.~\ref{fig:comparison}(\textbf{b})] and that opening a gap in Dirac points in 2D through symmetry-breaking generates a half-quantum concentration of Berry curvature near the avoided crossing [Fig.~\ref{fig:comparison}(\textbf{c})]. 
What are the analogous minimal models for solid-state realizations of a `Kalb-Ramond monopole' (more commonly called a \emph{tensor monopole} in the literature)? 
Since a minimum of three bands are necessary to generate the $\mathcal{DD}$ invariant in $PT$-symmetric models, it seems natural to start searching for strong concentrations of higher-form curvature (and of an unusually strong circular shift current response) in antiferromagnetic crystals hosting three-fold band degeneracies [Fig.~\ref{fig:comparison}(\textbf{d})]. 
Such \emph{triple nodal points}~\cite{Zhu:2016}, which have been classified across all magnetic space groups~\cite{Lenggenhager:2022}, should therefore be revisited from this newly accessed vantage point.
Note also that elementary \emph{chiral-symmetric} models of a tensor monopole have been theoretically described~\cite{Palumbo:2018} and experimentally simulated in a range of synthetic platforms~\cite{Tan:2021,Weisbrich:2021,Chen:2022,Zhang:2024,Mo:2025}, though they do not directly translate to realistic solid-state Hamiltonians without chiral symmetry.

In a similar spirit, higher-fold band degeneracies dubbed multifold fermions~\cite{Bradlyn:2016} could provide means for realizing the tensor monopole in $k\cdot p$ models with four or more energy bands.
Such construction has already been pointed out at the theoretical level for the case of the fourfold degenerate spin-$3/2$ fermions~\cite{Zhu:2020}.
In this context, note that non-linear photoconductivities of certain multifold fermions have been investigated~\cite{Ni:2020,Hsu:2023}; however, reports of a quantized shift-current response in these computations are presently missing.
The principal obstacle in facilitating such endeavors is the lack of a general understanding of the symmetry conditions that admit realizations of the $3$-form curvature and of a quantum of the $\mathcal{DD}$ invariant.
In particular, the construction of the Kalb-Ramond field in Eq.~(\ref{eqn:Kalb-Ramond-B}) has been cleanly formulated only for a narrow range of symmetry settings, such as Hamiltonians with chiral symmetry~\cite{Palumbo:2018}, spinless-$PT$ symmetry~\cite{Jankowski:2025}, or both~\cite{Palumbo:2021}.
The existing literature in this direction has not yet clarified how (if at all) one can extend these considerations to other Altland-Zirnbauer symmetry classes, such as those with spinful or spinless time-reversal symmetry, which are ubiquitous in crystals. 
In fact, even the elementary question of how the $3$-form curvature transforms under space group symmetries has not been directly addressed.

It is tempting to consider delicate and multigap topological models as toy laboratories for extending quantum geometric bounds and higher-form topology to further symmetry setups.
Indeed, while their unstable topology is typically lost upon the inclusion of additional bands, the $\mathcal{DD}$ invariant can remain well-defined after such a modification and may even enter the response theory, as exemplified by Eq.~(\ref{eqn:circular-photocurrent}).
However, while plentiful examples of unstable topologies are known in spinless-$PT$-symmetric and time-reversal-breaking models, very few extensions have, to date, been proposed in the presence of additional space group symmetries and in other Altland-Zirnbauer classes.
Besides the already mentioned RTP insulators~\cite{Nelson:2021} and chiral-symmetric insulators~\cite{Palumbo:2019}, the known examples are largely limited to Chern dartboard insulators~\cite{Brouwer:2023,Chen:2024}, spin Hopf insulators~\cite{Zhu:2023}, and delicate Wannier insulators~\cite{Guba:2025}.\footnote{The topology of the last two examples is, in fact, even more subtle, as it is protected by the closing of the Wannier gap rather than of the energy gap.}
A systematic classification of delicate and multigap topology across a broader range of settings, especially to symmetry setups not captured by homotopy groups, could provide a roadmap towards extending the higher-form topological description of energy bands, potentially unearthing geometrical or topological underpinning of further response coefficients.

Further opportunities for improvement exist on the computational front.
Specifically, any high-throughput search for a solid-state realization of the tensor monopole could be significantly accelerated by a reformulation of the curvature in Eq.~(\ref{eqn:3-form-curvature}) in a manner that accounts for the topological nature of its integral. 
Such a formulation is well recognized for the Chern number in 2D by virtue of the Fukui-Hatsugai-Suzuki algorithm~\cite{Fukui:2005}, which has been recently generalized to the winding number of unitary matrices in 3D~\cite{Shiozaki:2024}.
Specifically for the chiral-symmetric Hamiltonians in Eq.~(\ref{eqn:chiral-model}), a relation between the $\mathcal{DD}$ invariant and a three-dimensional winding number has been identified, implying that at least this particular realization of higher-form topology is potentially amenable to the computational approach of Ref.~\citenum{Shiozaki:2024}.
However, the question of whether this method may be further extended to also study the $\mathcal{DD}$ invariant in spinless-$PT$-symmetric systems, including prospective crystalline compounds, has not been properly addressed.
In this regard, note that the formulation of the Kalb-Ramond field in spinless-$PT$-symmetric models in Sec.~\ref{eqn:higher-forms} requires not only gauge fixing but also the construction of a smooth family of paths inside the momentum space; nevertheless, the resulting Kalb-Ramond field strength can be interpreted as a suitably anti-symmetrized interband torsion~\cite{Jankowski:2025}.
The latter quantity has been recently formulated in terms of gauge-invariant Bloch projectors\footnote{For example, the Hermitian connection~\cite{Ahn:2022} (mentioned in footnote~\ref{foot:C-connection}) is replaced in Ref.~\citenum{Mitscherling:2025} by the \emph{quantum geometric connection} 
\begin{equation}
\mathcal{C}^{mn}_{abc} = \textrm{tr}[P_n \partial_b P_m\left(\partial_a \partial_c P_n + \partial_a P_m \partial_c P_n\right)],    
\end{equation}
where $P_n = \ket{u_n}\!\!\bra{u_n}$ is the gauge invariant projector onto band $n$.
The interband torsion and the Kalb-Ramond field strength arising in spinless-$PT$-symmetric systems can be obtained from antisymmetrization of $\mathcal{C}^{mn}_{abc}$ in the subscript indices. The issue of gauge ambiguity has also been addressed in this context through the formalism of \emph{convergent $r$-matrix} method~\cite{Song:2026}.}~\cite{Mitscherling:2025}, suggesting a possible route towards more efficient computations.
The identification of suitable material candidates through computational means would, in turn, open the stage for their ARPES and photoconductivity studies.

It should further be interesting to consider whether other types (including higher orders) of non-linear conductivity similarly support a deeper and hitherto overlooked topological formulations. 
For example, Ref.~\citenum{Ahn:2022} showed how various third-order photoconductivity tensors relate to quantities derived from the QGT; however, their prospective topological reformulations remains unknown.
In addition, while this Perspective ties the concept of quantum geometry and higher-form topology by virtue of optical responses of electronic systems, such measurements are not applicable to classical simulators with charge-neutral excitations.
It thus remains to identify further smoking-gun experimental fingerprints of QGT and higher-form topology that could meaningfully be probed in such synthetic platforms.

It is also natural to wonder if higher-form topology is limited to `pristine' single-particle Hamiltonians or whether it survives generalization to the `dirty' cases with disorder and strong correlations.
In this respect, it is worth recalling the extension of the Chern number by Niu and Thouless to such a more general scenario by virtue of twisted boundary conditions~\cite{Niu:1984}, where the role of momentum is replaced by the twist angles.
This idea naturally extends to the $\mathcal{DD}$ invariant~\cite{Jankowski:2025}, though it remains an open problem to find a concrete variational many-body wave function that realizes a nontrivial value of the invariant.
In addition, a connection to strong disorder readily exists through the relation of multi-gap topology to ultra-localized topological insulators~\cite{Lapierre:2021,Lapierre:2024}.
It is interesting to speculate if this relation could be extended towards higher-form topology and non-linear responses.
Finally, let me point out that the language of higher-form topological structures has, over the past two years, been extended to matrix-product states (i.e., quantum many-body states with short-range entanglement) in 1D~\cite{Ohyama:2024,Shiozaki:2025,Sommer:2025,Ohyama:2025} and even to projected entangled pair states in 2D~\cite{Ohyama:2025b}.
In addition, several works have also opened the door to consider realizations of higher-form topology and tensor monopoles in non-Hermitian matter~\cite{Zhu:2021,Yang:2025}.

Overall, the early stage of research into higher-form topology in crystalline matter  (alongside the associated questions into quantum geometry and non-linear optics) suggests that impactful contributions can be achieved across all segments of the discovery process, ranging from mathematical characterization and space-group symmetry analysis, through toy models and materials design, to experimental investigations.
In concert, these capabilities can deliver definitive tests of quantum-geometry-based bounds on non-linear optical responses, including topological principles for quantized shift currents and a potential route toward a new class of photovoltaic materials.

\section{Acknowledgments}

A version of this text has previously appeared as a 
\emph{Scientific Perspective} in the newsletter of the Swiss MaNEP (Materials with Novel Electronic Properties) network~\cite{Bzdusek:2025}.
I would like to thank Frédéric Mila for inviting me to contribute to MaNEP with a short perspective on recent progress in topological band theory.
I am further grateful to Aris Alexandradinata, Titus Neupert, Robert-Jan Slager, and Bohm-Jung Yang for their valuable comments on an early version of this Perspective, to Niclas Heinsdorf for his exposition on gerbes in the context of matrix-product states, and to Giandomenico Palumbo for his insightful comments on the recent studies of bundle gerbes in band structures.
Finally, I would like to express my special gratitude to Wojciech Jankowski for his critical reading and extensive feedback on the entire text. 
This work was supported by Starting Grant No.~211310 from the Swiss National
Science Foundation.

\bibliography{references}

\end{document}